\documentclass[twocolumn,prl,showpacs,preprintnumbers,superscriptaddress]{revtex4-2}
\usepackage{amsmath}
\usepackage{dcolumn}
\usepackage{bm}
\usepackage{epsfig}
\usepackage{epstopdf}
\usepackage{graphicx}
\usepackage{amsfonts}
\usepackage{amssymb}
\usepackage{mathrsfs}
\usepackage{color}
\usepackage{appendix}
\usepackage{multirow}
\usepackage{graphicx}
\usepackage{afterpage}
\usepackage{rotating}
\usepackage{physics} 
\usepackage{siunitx}
\usepackage{natbib}
\usepackage{adjustbox}
\usepackage{svg}
\usepackage{orcidlink}
\usepackage{xcolor}
\usepackage{array}
\usepackage{hyperref} 
\hypersetup{colorlinks=true, linkcolor=blue, citecolor=blue, urlcolor=blue} 
\usepackage{makecell}  
\usepackage[normalem]{ulem}
\usepackage{soul}
\usepackage{booktabs}
\sethlcolor{cyan}
\setcellgapes{6pt}
\makegapedcells

\begin{document}

	\title{A Liquid-Nitrogen-Cooled $^{40}\mathrm{Ca}^{+}$ Ion Optical Clock with a Systematic Uncertainty of $4.4 \times 10^{-19}$}

	\author{Baolin Zhang}
	\thanks{These authors contributed equally to this work.}
	\affiliation{State Key Laboratory of Magnetic Resonance and Atomic and Molecular Physics, Innovation Academy for Precision Measurement Science and Technology, Chinese Academy of Sciences, Wuhan 430071, China}
	
	\author{Zixiao Ma}
	\thanks{These authors contributed equally to this work.}
	\affiliation{State Key Laboratory of Magnetic Resonance and Atomic and Molecular Physics, Innovation Academy for Precision Measurement Science and Technology, Chinese Academy of Sciences, Wuhan 430071, China}
	\affiliation{University of Chinese Academy of Sciences, Beijing 100049, China}
	
	\author{Yao Huang}
	\email{yaohuang@apm.ac.cn}
	\affiliation{State Key Laboratory of Magnetic Resonance and Atomic and Molecular Physics, Innovation Academy for Precision Measurement Science and Technology, Chinese Academy of Sciences, Wuhan 430071, China}
	
	\author{Huili Han}
	\affiliation{State Key Laboratory of Magnetic Resonance and Atomic and Molecular Physics, Innovation Academy for Precision Measurement Science and Technology, Chinese Academy of Sciences, Wuhan 430071, China}
	
	\author{Ruming Hu}
	\affiliation{State Key Laboratory of Magnetic Resonance and Atomic and Molecular Physics, Innovation Academy for Precision Measurement Science and Technology, Chinese Academy of Sciences, Wuhan 430071, China}
	\affiliation{University of Chinese Academy of Sciences, Beijing 100049, China}
	
	\author{Yuzhuo Wang}
	\affiliation{State Key Laboratory of Magnetic Resonance and Atomic and Molecular Physics, Innovation Academy for Precision Measurement Science and Technology, Chinese Academy of Sciences, Wuhan 430071, China}
	\affiliation{University of Chinese Academy of Sciences, Beijing 100049, China}
	
	\author{Huaqing Zhang}
	\affiliation{State Key Laboratory of Magnetic Resonance and Atomic and Molecular Physics, Innovation Academy for Precision Measurement Science and Technology, Chinese Academy of Sciences, Wuhan 430071, China}
	
	\author{Liyan Tang}
	\affiliation{State Key Laboratory of Magnetic Resonance and Atomic and Molecular Physics, Innovation Academy for Precision Measurement Science and Technology, Chinese Academy of Sciences, Wuhan 430071, China}
	
	\author{Tingyun Shi}
	\affiliation{State Key Laboratory of Magnetic Resonance and Atomic and Molecular Physics, Innovation Academy for Precision Measurement Science and Technology, Chinese Academy of Sciences, Wuhan 430071, China}
	
	\author{Hua Guan}
	\email{guanhua@apm.ac.cn}
	\affiliation{State Key Laboratory of Magnetic Resonance and Atomic and Molecular Physics, Innovation Academy for Precision Measurement Science and Technology, Chinese Academy of Sciences, Wuhan 430071, China}
	\affiliation{Hefei National Laboratory, University of Science and Technology of China, Hefei, 230088, China}
	\affiliation{Wuhan Institute of Quantum Technology, Wuhan 430206, China}
	
	\author{Kelin Gao}
	\email{klgao@apm.ac.cn}
	\affiliation{State Key Laboratory of Magnetic Resonance and Atomic and Molecular Physics, Innovation Academy for Precision Measurement Science and Technology, Chinese Academy of Sciences, Wuhan 430071, China}
	\affiliation{Hefei National Laboratory, University of Science and Technology of China, Hefei, 230088, China}
	
	\date{\today}
	
	\begin{abstract}
		
		We report a single-ion optical clock based on the $4\mathrm{S}_{1/2} \rightarrow 3\mathrm{D}_{5/2}$ transition of the $^{40}\mathrm{Ca}^+$ ion, operated in a liquid nitrogen cryogenic environment, achieving a total systematic uncertainty of $4.4 \times 10^{-19}$. We employ a refined temperature evaluation scheme to reduce the frequency uncertainty due to blackbody radiation (BBR), and 3D sideband cooling(SBC) to minimize the second-order Doppler shift. We have precisely determined the average Zeeman coefficient of the $^{40}\mathrm{Ca}^+$ clock transition to be $14.345(15)~\mathrm{Hz}/\mathrm{mT}^2$, thereby significantly reducing the quadratic Zeeman shift uncertainty. Moreover, the cryogenic environment enables the lowest reported heating rate due to ambient electric field noise in trapped-ion optical clocks.

	\end{abstract}
	
	\maketitle
		
	\textit{Introduction—}Optical frequency is currently the physical quantity with the highest measurement precision. Over the past two decades, with the advancement of atomic and optical theories and technologies, optical clocks have undergone substantial improvements, 
	such as laser cooling~\cite{Chen2017}, ultra-stable laser~\cite{Hafner2015,Oelker2019}, precise evaluation of the BBR shift~\cite{Steinel2023, Arnold2018, Brewer2019, Hassan2025}, and other frequency shift suppression methods~\cite{Huntemann2012,Dube201402,Arnold2020,Steinel2023,Lange2020,Chou2010}. At present, many optical clocks have achieved uncertainties at the order of ~$10^{-18}$ or even $10^{-19}$~\cite{Aeppli2024,marshall2025,Brewer2019,Ushijima2015,Huang2022,Tofful2024,Dorscher2021,zhiqiang2023,Li2024}, enabling optical clocks to play an irreplaceable role in the field of precision measurement, such as investigating whether the fine-structure constant $\alpha$ varies over time~\cite{Godun2014}, testing Lorentz invariance~\cite{Lange2021,Sanner2019}, and detecting gravitational waves and dark matter\cite{Kolkowitz2016,Filzinger2023}. Meanwhile, the ultra-high precision of optical clocks is driving the redefinition of the SI second~\cite{Dimarcq_2024} and has further promoted their applications in navigation and geodesy~\cite{takamoto2020,McGrew2018,meh2018}.
	
	The $^{40}\mathrm{Ca}^{+}$ ion exhibits a simplified energy level structure and demands fewer lasers operating at significantly reduced power—typically several hundred microwatts—when compared to other optical clock systems such as Sr~\cite{Aeppli2024}, Yb~\cite{McGrew2018}, Yb$^{+}$~\cite{Dorscher2021} and Al$^{+}$~\cite{marshall2025}. This streamlined configuration enhances experimental feasibility while maintaining high performance~\cite{Zhangbaolin2020}. Moreover, it has a magic trapping RF (radio frequency) that one can cancel the micromotion-induced Doppler and Stark shifts, making it one of the  ideal optical clock candidates approaching very low total systematic uncertainty\cite{Dube201402,Huang2019}. The main contributor to the systematic uncertainty of the $^{40}\mathrm{Ca}^+$ ion optical clock is the BBR shift. To address this, we previously utilized a liquid-nitrogen cooling scheme to reduce the  BBR shift uncertainty to $2.7 \times 10^{-18}$\cite{Huang2022}. However, due to the accuracy of the temperature evaluation, the BBR shift remains the largest contributor to the total systematic uncertainty~\cite{Huang2022,Zengmengyan2023}. Additionally, the second-order Doppler shift arising from secular motion is also one of the main factors restricting the further reduction of the total systematic uncertainty of the $^{40}\mathrm{Ca}^+$ ion optical clock to the $10^{-19}$ level~\cite{Huang2022}.
	
	In this Letter, we report on the research progress of the second-generation liquid-nitrogen cryogenic ion optical clock \(\mathrm{Ca}^{+}\!\text{-}2\) (LNCIOC \(\mathrm{Ca}^{+}\!\text{-}2\)), which has achieved a total systematic uncertainty of $4.4 \times 10^{-19}$. This is mainly attributed to our improvements in the design and installation scheme of the BBR cavity and ion trap, the incorporation of 3D SBC, and the precise measurement of the quadratic Zeeman coefficient for the $^{40}\mathrm{Ca}^+$ ion's clock transition.
	
	The setup is illustrated in Fig.~S1(a), and details of the trap device can be found in references~\cite{Chou2010, Huang2022}. The trap RF frequency is set at 24.801~MHz, which is close to the magic trapping frequency~\cite{Dube2014,Huang2019}. A DC voltage of 160~V is applied to both endcap electrodes. To separate the two secular motion frequencies in radial directions thus enabling effective SBC for all 3 motional modes, we apply DC voltages of $\pm8$~V to the two RF coupled electrodes. Additionally, RF amplitude stabilization is introduced to stabilize the radial secular motion frequencies, with a long-term drift of less than 2~kHz~\cite{ma2025}. The excess micromotion (EMM) of the ion is evaluated through the sideband spectra observed by three mutually orthogonal clock laser beams (beam~1, 2, and~3 in Fig.~S1(b)) and is suppressed by optimizing the voltages applied to the compensation electrodes. The vacuum system of the liquid-nitrogen cryogenic optical clock is surrounded by four layers of magnetic shielding to reduce the magnetic field noise\cite{Zeng2023}. All lasers used in LNCIOC \(\mathrm{Ca}^{+}\!\text{-}2\) are shown in Fig.~S1(b). The quantization axis is along the $y$ axis direction with a magnetic field of approximately 37.2~$\mu$T.
	
	The sequence starts with a 1.5~ms Doppler cooling pulse. Then the $^{40}\mathrm{Ca}^+$ ion is optically pumped into one of the two sub-levels of the $\mathrm{S}_{1/2}$ state using the 397~nm $\sigma^+/\sigma^-$ and 8667~nm lasers, achieving a state-preparation efficiency exceeding 99\% within a 5-$\mu$s pulse length. Subsequently, the $^{40}\mathrm{Ca}^+$ ion is further cooled to near the 3D motional ground state through resolved SBC (10~ms). To avoid the AC Stark shift, all lasers except the clock laser are blocked using mechanical shutters during the clock interrogation. The clock transition is probed with a hyper Ramsey scheme, employing a $\pi$-pulse duration of 4~ms and a free evolution time of 80~ms ~\cite{Huntemann2012,Yudin2010}. Finally, a 2-ms-long state detection pulse is performed to determine the ion’s final state.
	
	During the clock operation, we alternately lock the clock laser onto three pairs of Zeeman transitions: $\left| \mathrm{S}_{1/2}, m = \mp 1/2 \right\rangle \rightarrow \left| \mathrm{D}_{5/2}, 
	m = \mp 3/2 \right\rangle$, $\left| \mathrm{S}_{1/2}, m = \pm 1/2 \right\rangle \rightarrow \left| \mathrm{D}_{5/2}, m = \mp 1/2 \right\rangle$, and 
	$\left| \mathrm{S}_{1/2}, m = \mp 1/2 \right\rangle \rightarrow \left| \mathrm{D}_{5/2}, m = \mp 5/2 \right\rangle$, to cancel out the first-order Zeeman shift, the electric quadrupole shift, and the tensor Stark shift~\cite{Dube2014,Huang2019}.

	\textit{Blackbody radiation shift—}The BBR shift can be expressed as~\cite{Farley1981,Porsev2006}:
	\begin{equation}
		\Delta \nu_{\mathrm{BBR}} = -\frac{1}{2} \left(831.9~\mathrm{V\,m^{-1}}\right)^2 \left(\frac{T(\mathrm{K})}{300}\right)^4 \Delta \alpha_0 \left(1 + \eta\right),
	\end{equation}
	where $T$ is the ambient temperature, $\Delta \alpha_0$ is the differential static polarizability for the $^{40}\mathrm{Ca}^+$ ion's clock transition, and $\eta$ represents a minor dynamic correction that arises from the energy-level structure of the $^{40}\mathrm{Ca}^+$ ion and the spectral distribution of the BBR field~\cite{Porsev2006}. Compared to room-temperature optical clocks, the liquid-nitrogen cryogenic environment reduces the ambient temperature by approximately a factor of 4, leading to a reduction of the BBR shift uncertainty by more than 200 times with a similar evaluated temperature uncertainty. Consequently, the uncertainty in the BBR shift caused by $\Delta \alpha_0$ and $\eta$ is on the order of $10^{-21}$.
	
	Compared to the first-generation liquid-nitrogen cryogenic $^{40}\mathrm{Ca}^+$ ion optical clock, we have made several key improvements in the design and assembly of the BBR cavity, ion trap, and electrodes. As shown in Fig.~S1, the ventilation holes in the BBR cavity---originally included to maintain vacuum consistency with the external chamber---are now sealed with oxygen-free copper covers to effectively block thermal radiation from the room-temperature environment. Additionally, to enhance temperature uniformity among the ion trap, ion trap holder, endcap electrodes, BBR cavity, and  bottom of the Liquid-Nitrogen Cryogenic container(LNC), thin silver foils are inserted at the mechanical interfaces to enhance thermal contact and ensure their temperatures closely track that of the LNC. Furthermore, phosphor-bronze wires are employed as the conductors for both temperature sensors and electrodes. Before entering the BBR cavity, these wires are wound around the LNC several times to ensure their temperatures fully equilibrate with that of the LNC. Collectively, these modifications significantly enhance the temperature uniformity of the ambient environment.
	
	\begin{figure}[!htb]
		\centering
		\includegraphics[scale=0.62]{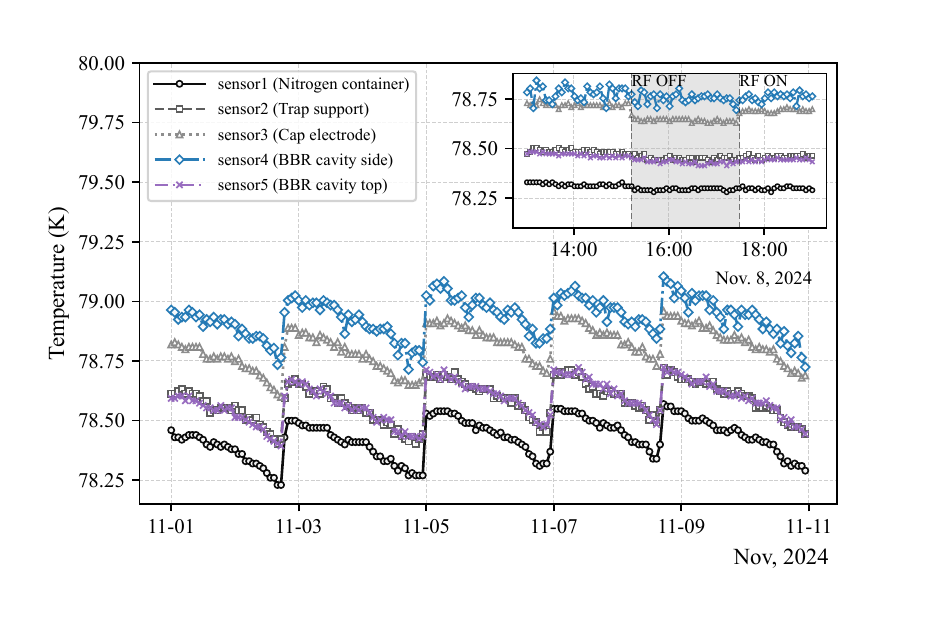}
		\caption{\small{Temperature measurements of the LNCIOC \(\mathrm{Ca}^{+}\!\text{-}2\) ion optical clock~2. The abrupt temperature increases are due to the periodic refilling of liquid nitrogen.}}
		\label{figure1}
	\end{figure}

	We used PT100 thermistors in a three-wire configuration to eliminate temperature measurement errors from wire resistance. All sensors were individually calibrated to an accuracy of 0.1~K over the 75–85~K temperature range. In LNCIOC \(\mathrm{Ca}^{+}\!\text{-}2\), five sensors were employed to monitor the temperatures at the bottom of the LNC, the endcap electrodes, the ion trap holder, and the BBR cavity. See Fig.~S1(c) in the supplementary material for the detailed locations of these components.
	
	To directly assess the temperature of the ion trap, we constructed a replica system with an identical ion-trapping setup and BBR cavity to that of LNCIOC \(\mathrm{Ca}^{+}\!\text{-}2\). An additional temperature sensor was attached to the ion trap within this replica system. The ion trap exhibits the highest temperature among all components, due to its direct connection to four electrodes. When the RF is switched on or off, a temperature change of approximately 0.4~K is observed (see Fig.~S2). We took the upper and lower bounds of all temperature sensors---spanning both the operational and replica systems---yielding a temperature estimate of $79.5 \pm 1.4$~K (see Fig.~\ref{figure1} and Fig.~S2). Considering the 0.1~K calibration uncertainty, the final BBR temperature is evaluated as $79.5 \pm 1.5$~K, leading to a BBR uncertainty of $3.5 \times 10^{-19}$, nearly an order of magnitude lower compared to the first-generation liquid-nitrogen cryogenic $^{40}\mathrm{Ca}^+$ optical clock.

	\textit{Second-order Doppler shift—}We implemented 3D SBC pulses in LNCIOC \(\mathrm{Ca}^{+}\!\text{-}2\) to suppress the second-order Doppler shift. Due to small Lamb-Dicke parameters for the 3D secular motions, which are $\eta_x = 0.025$, $\eta_y = 0.027$, and $\eta_z = 0.043$ respectively, the second-order sidebands are too weak to be experimentally resolved~\cite{wineland1998}. Therefore, we sequentially selected three first-order red sidebands to reduce the occupation numbers. Detailed sequences are illustrated in Fig.~S3.
	
	\begin{figure}[!htb]
		\centering
		\includegraphics[scale=0.55]{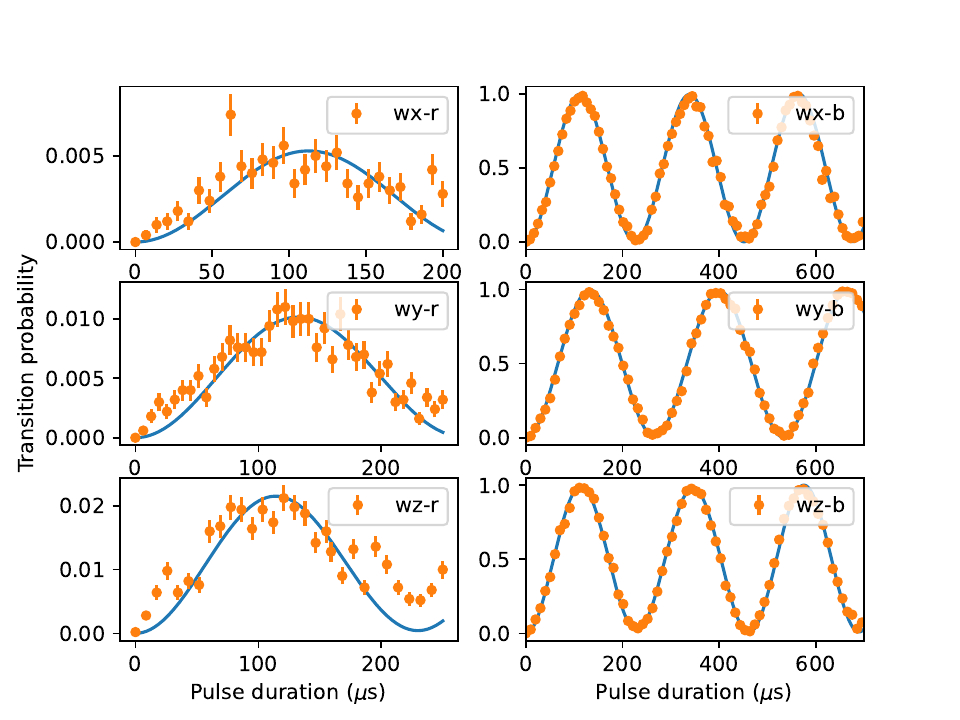}
		\caption{\small{Red- and blue-sideband Rabi oscillations for the three motional modes after SBC. The orange dots represent the experimentally measured transition probabilities, and the blue solid lines are thermal-distribution fits to these data.}}
		\label{figure2}
	\end{figure}
	
\begin{figure}[!htb]
		\centering
		\includegraphics[scale=0.55]{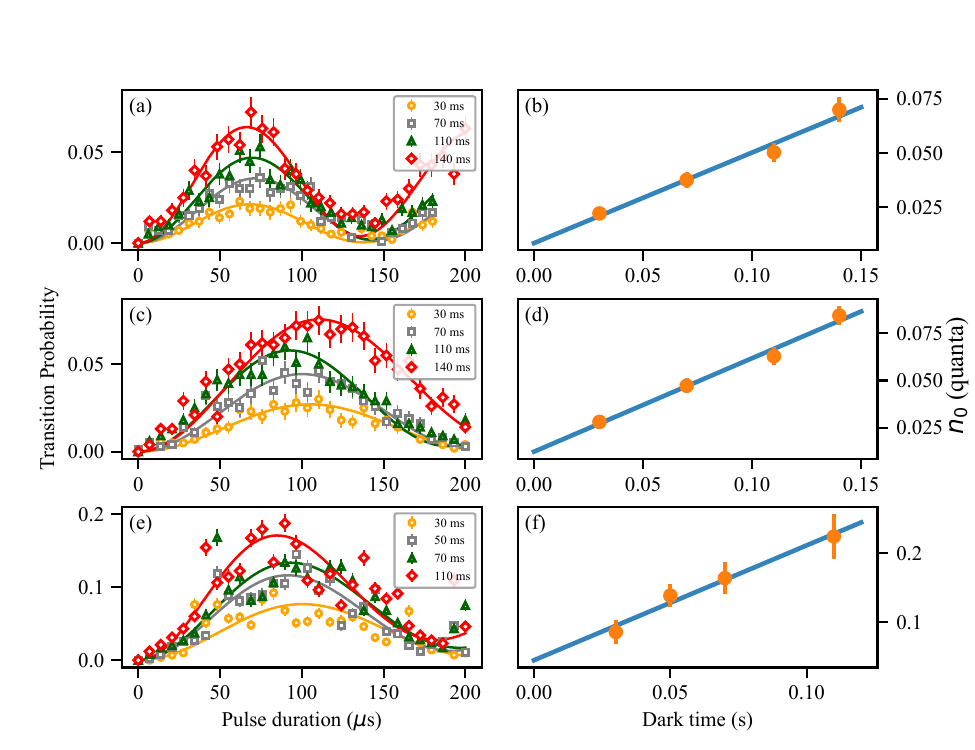}
		\caption{\small{Measurement of the heating rates.Panels~(a), (c), and (e) present red-sideband Rabi oscillations of the $X$-, $Y$-, and $Z$-motional modes, respectively, measured at different dark times following SBC. By fitting these curves with thermal-distribution models, the mean occupation numbers at different delays are extracted. (b), (d), and (f) show linear fits to the data in (a), (c), and (e), respectively. The slope of each fit corresponds to the heating rate, and the intercept gives the mean occupation number after SBC. All fit results include 95\% confidence intervals as uncertainty estimates.}}
		\label{figure3}
	\end{figure}

	Results of 3D SBC are shown in Fig.~\ref{figure2}. Assuming that the Fock-state distribution is thermal after SBC, we extract the mean occupation number, $\bar{n}$, by fitting the experimental data of the red-sideband Rabi oscillation. Furthermore, as shown in Fig.~\ref{figure3}, by inserting a 'dark time' ranging from 30 to 140~ms between SBC and red-sideband interrogation, we extracted the heating rates $\dot{\bar{n}}$ due to ambient electric field noise. Under our experimental parameters, the hypothesis of a thermal distribution is validated by the simulation process of SBC (see Supplementary Materials), despite being inconsistent with the results observed in Raman SBC~\cite{Chen2017,Rasmusson2021}. In addition to directly fitting the red sideband rabi oscillation after SBC, we also utilize the $y$-intercepts in Figs.~\ref{figure3}(b), \ref{figure3}(d), and \ref{figure3}(f) to evaluate the average occupation number $n_0$ after SBC. The final assessment of $n_0$ incorporates both of the two approaches.

	\begin{table}[htbp]
		\caption{Evaluation of the second-order Doppler shift for LNCIOC \(\mathrm{Ca}^{+}\!\text{-}2\).}
		\begin{tabular*}{\hsize}{@{\extracolsep{\fill}}lccc}
			\toprule
			& $x$ & $y$ & $z$ \\
			\midrule
			$\omega / 2\pi$ (MHz) & 3.531 & 3.166 & 2.580 \\
			$\eta$ & 0.025 & 0.027 & 0.043 \\
			TD/quantum ($10^{-19}$) & 3.9 & 3.5 & 2.9 \\
			$n_0$ (quanta) & 0--0.03 & 0--0.04 & 0--0.11 \\
			$\dot{\bar{n}}$ (quanta/s) & 0.43(7) & 0.53(12) & 1.31(43) \\
			\midrule
			\textbf{Total} [$10^{-19}$] & \multicolumn{3}{c}{$-5.8 \pm 0.4$} \\
			\bottomrule
		\end{tabular*}
		\label{tab:2order_doppler}
	\end{table}
	
	Over a period of nearly one month, we carried out five repeated measurements of both the initial mean occupation numbers after SBC and the heating rates, as presented in Fig.~S4. Under an 80-ms interrogation time, the uncertainty in the second-order Doppler shift is evaluated to be $4 \times 10^{-20}$, as shown in Table~\ref{tab:2order_doppler}.

	\textit{Second-order Zeeman shift—}The output frequency of the $^{40}\mathrm{Ca}^+$ ion optical clock is determined by averaging over the three pairs of Zeeman transitions mentioned above. By comparing the output frequencies of room-temperature $^{40}\mathrm{Ca}^+$ ion optical clocks, we precisely measured the average second-order Zeeman coefficient $\alpha_z$ of the three pairs of Zeeman transitions. One clock served as a frequency reference, with its DC magnetic field fixed at a constant value $B_0$, and another served as the modulated clock with varing dc magnetic fields B. The frequency difference $\delta_z$ between the two clocks was measured as a function of $B$ and $B_0$, and can be described by the relation~\cite{Godun2014}:
	\begin{equation}
		\delta_z = \alpha_z (B^2 - B_0^2).
		\label{eq:zeeman_shift}
	\end{equation}
	
	By varying the DC magnetic field $B$ of the modulated clock, we obtained a set of frequency differences between the two clocks, as shown in Fig.~S9. A linear fit to these data yielded $\alpha_z = 14.345(15)~\mathrm{Hz/mT}^2$ using the Monte Carlo method.
	
	We derived the DC magnetic field of LNCIOC \(\mathrm{Ca}^{+}\!\text{-}2\) from the measured Zeeman splittings, determining it to be $37.197(1)\ \mu\mathrm{T}$, corresponding to a second-order Zeeman shift uncertainty of $0.5 \times 10^{-19}$. Although the DC magnetic field may exhibit significant drift over a long period, daily variations remain well within this uncertainty budget.
	
	Furthermore, due to imperfections in the trap symmetry, the RF field generates an oscillating magnetic field at the ion's position, with a frequency equal to that of the RF~\cite{Gan2018}. This RF-induced AC magnetic field can be characterized by measuring its effect on the ratio of Landé $g$-factors associated with the clock transition~\cite{Ma2024,Arnold2019}. The root mean square of the $y$-component of the AC magnetic field were measured to be $ \langle B_y^2 \rangle = 11(9) \times 10^{-14}~\mathrm{T}^2.$ This yields an estimated upper bound of $7 \times 10^{-21}$ for the second-order Zeeman shift.

	\textit{Excess micromotion (EMM) shift—}There exists a magic drive frequency of $24.801(2)\ \mathrm{MHz}$ for the $^{40}\mathrm{Ca}^+$ ion clock transition, at which the EMM-induced second-order Doppler shift and the Stark shift cancel out~\cite{Dube2014,Berkeland1998}. The RF frequency was set to $24.801\ \mathrm{MHz}$ for LNCIOC \(\mathrm{Ca}^{+}\!\text{-}2\) to minimize the EMM shift. Through long-term measurements of the sideband spectra in three mutually perpendicular directions, the frequency shift caused by EMM is evaluated to be $(0 \pm 1.4) \times 10^{-19}$.

	\textit{Background gas collisions—}The collision-induced frequency shift (CFS) arising from interactions with background helium gas is evaluated using a nonperturbative quantum channel framework developed by A.~C.~Vutha ~\cite{Vutha2017}. This method has been validated in prior studies involving Sr$^+$~\cite{Vutha2017} and Al$^+$ ~\cite{Hankin2019,Davis2019} ions colliding with He and H$_2$. As it is not feasible to measure the pressure inside the BBR cavity, we use the pressure measured in the ambient room-temperature environment outside the BBR cavity, which is \(3 \times 10^{-9}~\mathrm{Pa}\), as an overestimation of the pressure within the BBR cavity. For the background gas at a temperature of \(78.5~\mathrm{K}\), a particular calculation approach yields a collision frequency shift of \(1.8 \times 10^{-19}\). Details can be found in the Supplementary Material.
	
	\textit{Other shifts—}The hyper Ramsey excitation scheme was employed to suppress the AC Stark shift induced by the clock laser~\cite{Huntemann2012, Huang2022} and the chirp effect induced by the AOM switching~\cite{Falke2012}, ensuring that both frequency shifts are less than $1 \times 10^{-19}$. Details can be found in the Supplemental Material. The total systematic frequency shifts of LNCIOC \(\mathrm{Ca}^{+}\!\text{-}2\) are presented in Table~\ref{tab:systematics}.

	\renewcommand{\arraystretch}{0.5}
	\begin{table}[htbp]
		\centering
		\caption{Systematic shifts and uncertainties of the $\mathrm{LNC} \cdot{ }^{40} \mathrm{Ca}^{+}$ion optical clock 2.}
		\begin{tabular}{lcc}
			\toprule
			\textbf{Effect} & \textbf{Shift ($10^{-19}$)} & \textbf{Uncertainty ($10^{-19}$)} \\
			\midrule
			BBR                          & 45.5  & 3.5  \\
			Excess micromotion           & 0.0   & 1.4  \\
			Secular motion               & -5.8  & 0.4  \\
			Quadratic Zeeman (DC)        & 482.9 & 0.5  \\
			Quadratic Zeeman (AC)        & 0.0   & $<0.1$ \\
			Background gas               & 0.0   & 1.8  \\
			Residual 1st Zeeman          & 0.0   & 0.2  \\
			Electric quadrupole          & 0.0   & 0.1  \\
			AOM phase chirp              & 0.0   & 1.0  \\
			Clock laser Stark            & 0.0   & $<0.1$  \\
			\midrule
			\textbf{Total}               & \textbf{522.6} & \textbf{4.4} \\
			\bottomrule
		\end{tabular}
		\label{tab:systematics}
	\end{table}

	\textit{Conclusion—}We have developed a second-generation liquid-nitrogen cryogenic \({}^{40}\mathrm{Ca}^{+}\) ion optical clock with an uncertainty of \(4.4 \times 10^{-19}\), which represents an improvement of approximately one order of magnitude compared to our previous result~\cite{Huang2022}. It is the lowest reported uncertainty among all optical clocks to date. The cryogenic environment not only significantly reduces the BBR shift and its uncertainty, but also markedly reduces the heating rates caused by ambient electric field noise. As the performance of ion optical clocks continue to improve, the heating effect has gradually become one of the main limiting factors for the system uncertainty and ion coherence time of optical clocks. Our research experimentally confirms the remarkable suppression effect of the cryogenic environment on the heating rate, thereby offering a viable approach for enhancing the performance of ion optical clocks.
	
	For LNCIOC \(\mathrm{Ca}^{+}\!\text{-}2\), the BBR shift remains the primary contributor to the systematic uncertainty. Developing \({}^{40}\mathrm{Ca}^{+}\) ion optical clocks cooled with helium at even lower temperatures or employing more accurate temperature assessment schemes holds promise for further reducing the BBR shift to the order of \(10^{-20}\). Meanwhile, by creating conditions to measure the ion exchange rate between two ions, one can precisely evaluate the vacuum level in directly cooled single-ion optical clocks. Additionally, developing a more comprehensive theoretical model for the collision frequency shift will enable further reduction of the background gas collision-induced frequency shift. In the next step, we will also enhance the performance of the clock laser and magnetic shielding to extend the clock interrogation time, aiming to achieve a \({}^{40}\mathrm{Ca}^{+}\) ion optical clock with a few \(10^{-16}/\sqrt{\tau}\) level stability. This, in turn, will allow the application of the \(10^{-19}\)-level precision of the \({}^{40}\mathrm{Ca}^{+}\) ion optical clock to precision measurements such as the frequency ratio between different optical clocks, the time variation of the fine-structure constant \(\alpha\), and the determination of millimeter-level height differences and tidal potentials.
		
	\textit{Acknowledgements—}We are deeply grateful to Bing Yan for providing the Ca\textsuperscript{+}–He potential data, and to A.~C.~Vutha for helpful discussions. This work is supported by the National Key R\&D Program of China (Grant Nos. 2022YFB3904001 and 2022YFB3904004), Innovation Program for Quantum Science and Technology (Grant No. 2021ZD0300901), the National Natural Science Foundation of China (Grant Nos. 12204494, 12121004, 12320101003, 11934014, 12374235 and 12174402), the CAS Youth Innovation Promotion Association (Grant No Y2022099), the Natural Science Foundation of Hubei Province (Grant Nos. 2022CFA013 and 2023EHA006), the CAS Project for Young Scientists in Basic Research (Grant Nos. YSBR-055 and YSBR-085), and Strategic Priority Research Program of the Chinese Academy of Sciences (XDB0920101 and XDB0920202).
	
	\bibliographystyle{apsrev4-2} 
	\bibliography{ref.bib}

\end{document}